# Radiation-induced Ionization Effects and Space Mission Requirements for Silicon Photonic Mach-Zehnder Modulators

Kellen P. Arnold[1], Joel B. Slaby[2], Nathaniel J. Karom[3], Anurag R. Veluri[4], C. Alex Kaylor[2], Andrew L. Sternberg[4], Dennis R. Ball[4], Ronald D. Schrimpf[1,4], Daniel M. Fleetwood[1,3,4], Stephen E. Ralph[2], Robert A. Reed[1,4], and Sharon M. Weiss[1, 3, 4]

[1.] Interdisciplinary Materials Science Graduate Program, Vanderbilt University, Nashville, TN 37235, USA
[2.] School of Electrical and Computer Engineering, Georgia Institute of Technology, Atlanta, GA 30332, USA
[3.] Department of Physics and Astronomy, Vanderbilt University, Nashville, TN 37235, USA
[4.] Department of Electrical and Computer Engineering, Vanderbilt University, Nashville, TN 37235, USA

**Abstract**
Photonic integrated circuits have become essential for meeting the growing global demand for high-capacity information processing and transport. Assessing their radiation tolerance is essential for deploying systems in radiation prone environments – including in space, high-energy particle accelerators, and defense radiation testing facilities – where the performance and compactness of photonic integrated circuits are increasingly advantageous. This work investigates the analog and digital radio frequency electro-optic performance of Mach-Zehnder modulators (MZMs) subject to 10-keV X-ray irradiation, which mimics cumulative ionization effects in space flight. Silicon photonic MZMs serve as excellent exemplars since they are interferometric devices comprised of elements common to many integrated photonic circuits. Under standard bias conditions, the irradiated MZMs exhibited significantly reduced bandwidth, a corresponding eye closure and baud rate dependent increases in the estimated error rate. The observed performance degradation is attributed to total ionizing dose effects which leads to hole trapping at the silicon/silicon dioxide waveguide interfaces as well as "fast traps" with energies near the conduction band edge. Notably, when MZMs were irradiated with all leads grounded, no radiation sensitivity to the electro-optic response was observed highlighting the importance of testing under standard operating conditions for ground-based radiation testing as well as on-orbit studies. Understanding the radiation induced performance degradation of MZMs and other integrated photonic devices is increasingly important for space and accelerator environments as performance requirements and deployment opportunities increase.

## Introduction

Silicon photonics has advanced steadily over recent decades, enabling the transition from research to commercial deployment. Today, seamless integration of photonic devices and circuits with traditional electronics has enabled unprecedented signal processing and transmission capacity.[1] Silicon photonics leverages the well-established silicon fabrication infrastructure that underpins modern electronics, permitting production of photonic integrated circuits (PICs) at commercial foundries for widespread adoption and co-packaging alongside traditional electronic ICs.[2,3]

As terrestrial deployment of photonics accelerates in commercial markets, the size, weight, and power advantages also make PICs of interest to the commercial space industry and the defense industrial base for communications (on-board, satellite-to-satellite, satellite-to-ground), sensing, LiDAR, encryption, and navigation.[4,5] Given the harsh environment in space, understanding the radiation response of integrated silicon photonic (SiPh) devices is critical.[6,7] However, relatively few studies have addressed radiation sensitivities of PIC components, which include waveguides,[8–11] modulators,[12–22] ring resonators and filters,[23,24] and photodiodes.[12,25–30]

Fundamental to the function of most integrated photonic systems is the modulator, which encodes electronic data into the amplitude and/or phase of a continuous-wave laser. The two primary configurations for on-chip photonic modulators are Mach-Zehnder modulators (MZMs) and microring modulators,[31] where both designs rely on modulation of the material refractive index by changing carrier density within the waveguide path. The MZM is an interferometric device supporting both amplitude and phase modulation and is therefore the workhorse of high-capacity optical communication systems. MZMs also offer a wide optical window, a large electrical bandwidth, and relatively low temperature sensitivity. The MZM is commonly deployed for both digital and analog radio frequency (RF) systems, making it an excellent candidate for communications applications in space. The microring modulator relies on an optically resonant ring acting as a switchable filter and is therefore primarily deployed as an amplitude modulator for digital systems. Both modulator types rely on p-n junctions to affect carrier density and include precise waveguide fabrication. The structure, modulation mechanism and the deployment opportunities make the MZM an optimum device to fully investigate and understand mechanisms that may impact PIC performance within a harsh radiation environment. Therefore, a thorough investigation is essential for reliable deployment in satellite and other space mission platforms. Although this work focuses on MZMs, the methodology and findings are applicable to microring modulators and emerging PIC designs,[32] where similar radiation degradation mechanisms can be expected.

The relatively few harsh radiation environment studies on SiPh MZMs have employed varied approaches. First, single event effects (SEEs), which occur when energetic particles track through active regions and deposit transient charge, have been assessed using pulsed laser irradiation to experimentally simulate heavy ion strikes. In general, single event transients (SETs) can affect both electrical and optical signals,[16,17] although most effects on optical properties occur at relatively high laser pulse energies, which correspond to heavy ions with high linear energy transfer in a deployed system on-orbit. Dominant SETs and upset modes for the space environment have been shown to occur in the modulator driver and trans-impedance amplifier of the receiver,[33] motivating development of radiation-tolerant modulator drivers.[22,34–37] Second, total ionizing dose (TID) effects, which induce charge trapping in dielectrics and formation of traps at semiconductor/oxide interfaces are known to affect carrier densities and dynamics in microelectronic devices, are studied using ionizing radiation sources (e.g., X-rays, gamma rays, protons).[38,39] Changes resulting from oxide and interface trap occupation often lead to degradation of threshold voltage and leakage currents. Impacts on SiPh MZM performance resulting from TID include changes in the static behavior of phase modulators in MZMs, which in turn affects modulation efficiency.[12,15,18–20,40] Zeiler, *et al.*[12] was the first to report that TID results in hole trapping that causes "pinch-off" in the carrier depletion phase modulator, leading to performance degradation that worsens when the MZM is biased during irradiation, as opposed to being grounded during irradiation. Third, displacement damage occurs when energetic particles (charged or neutral) collide with atoms within a crystalline lattice, causing vacancies or defect clusters to form.[41,42] Defect accumulation in sensitive device volumes leads to carrier lifetime degradation and defect-assisted non-radiative recombination pathways. In SiPh MZMs, no sensitivity was identified under 20-MeV neutron irradiation, suggesting that displacement damage is not a key contributor to cumulative radiation effects.[12,19] Two studies have been performed flying silicon photonic chips with MZMs in low Earth orbit at the International Space Station.[43,44] However, no on-orbit characterization was performed and these chips were likely electrically floating for the exposure duration.

While prior studies have reported how MZM performance is affected in the presence of high energy radiation, a clear understanding of the RF performance of an MZM in a harsh radiation environment remains incomplete. Han *et al.*[21] observed RF performance degradation modes in SiPh MZMs, identifying eye closure following 3-MeV proton irradiation, although the lowest TID levels studied exceed 500 Mrad($SiO_2$), which is beyond that encountered within the duration of space missions. Zhao *et al.*[45] reported analog and digital RF characteristics of SiPh MZMs under proton and gamma irradiation. While electro-optic bandwidth changes were reported in some devices under gamma irradiation (TID only), no changes were reported under 0.17- and 30-MeV proton irradiation (TID and displacement damage) where the proton irradiation TID levels exceeded those of gamma irradiation. In these experiments, each dose and fluence exposure was performed on a separate chip, which lends itself to large levels of uncertainty when comparing results from chip-to-chip. Considering all prior work, somewhat inconsistent conclusions have been drawn when evaluating SiPh MZMs in intense radiation environments, necessitating a unified test methodology that evaluates both the physical mechanisms of degradation under irradiation and impacts to high-speed performance of photonic devices.

In this work, we quantify TID effects on the RF performance of SiPh MZMs and identify a likely mechanism. Furthermore, we identify appropriate test conditions for radiation effects investigations and space mission reliability qualification of integrated photonic devices and circuits. Using 10-keV X-ray irradiation, the analog and digital RF electro-optic response of the SiPh MZM is measured as a function of TID. Measurements were performed *in situ*, but not in flux, meaning that the device is fully addressable during irradiation and kept in place throughout each test sequence, but all measurements are performed after the completion of each irradiation step to evaluate fixed TID levels.

Significant reduction in the electro-optic bandwidth, closure of the eye diagram, and baud rate dependent increases in the estimated error rate are observed for devices actively biased under standard operating conditions during irradiation. Importantly, for X-ray experiments where all leads are grounded during irradiation, no high-speed electro-optic response sensitivity is observed, emphasizing that test conditions play an important role in qualifying photonic parts for space missions and elucidating physical degradation modes in harsh environments. The underlying physical degradation mechanisms are attributed to charge collection near the waveguide/oxide interfaces, which is supported by experimental observation of carrier depletion phase modulation efficiency reduction, especially when biased during irradiation, and technology computer-aided design (TCAD) simulation. Moreover, frequency dependent alterations to carrier dynamics are observed in the TID-induced electro-optic bandwidth degradation trends, which are attributed to "fast traps" with charge capture and emission time constants in line with GHz frequencies.

**Device design**

SiPh MZMs for this study were created with the Fotonix™ process design kit (PDK) from GlobalFoundries (GF) and were configured with RF electrodes, electrical contact pads, and 36 Ω terminations. The test die (5 mm square tile) were fabricated by GF on 300 mm wafers from a multi-project wafer run;[46] an optical micrograph of the tile is shown in **Fig. 1a**. Optical input and output are achieved using diffractive grating couplers from the standard foundry PDK. The grating couplers and electrical bond pads are arranged to allow fiber connections without interfering with simultaneous probing and/or wire bonds and irradiation of the die. The MZM includes 2mm long high-speed p-n junction carrier depletion phase modulators (phase modulators) and low speed thermal phase tuners (thermal tuners) to ensure bias at the optical quadrature point. The phase modulators are electrically contacted using a ground-signal-ground-signal-ground (GSGSG) pad configuration with a common ground plane connected to the anode of both phase modulators and the signal pads connected to each modulator arm (we designate one arm as reference and one as signal). The thermal tuners are also indicated in the inset and the direct current (DC) bond pads of the active thermal tuner are indicated at the top of the layout away from the DUT and GSGSG contacts. The cross-sectional schematic view of the phase modulator is shown in **Fig. 1b**. A sub-200 nm silicon (Si) rib waveguide device layer (red/blue) is patterned atop a buried oxide ($SiO_2$, light blue). The n-type (n+ in blue, n++ in dark blue) and p-type (p+ in red, p++ in dark red) implant regions are indicated and the depletion region is illustrated as designed in the device layout, positioned at 1/4 the total waveguide width, which is 400 nm. A sub-100 nm silicon nitride (SiN) layer (purple) is embedded in oxide and positioned 25 nm above the surface of the silicon waveguide. A full metallization stack is included in the MZM design – only the anode (p++) and cathode (n++) contacts are shown (brown) and the optical mode is unaffected by all metal layers.

A photograph of a representative test package is shown in **Fig. 1c**, where the optical input and output are fiber attached and the DC and RF electrical connections to the chip are wire bonded, enabling a fully addressable handheld test package for *in situ* testing in the radiation source. Testing was performed using both an RF printed circuit board (PCB) design and RF probes. We note that the PCB design using ball bonds reduced the observed bandwidth compared to the RF probe. The PCB added approximately 10 dB loss at 25 GHz (vs. <2 dB at 50 GHz for the probe). However, the fully packaged assembly provides sufficient AC signal to measure RF performance, simplifies setup for irradiation, and provides certainty that consistent contact is maintained throughout the irradiation test procedure. Full details of the packaging technique can be found in the **Methods** section.

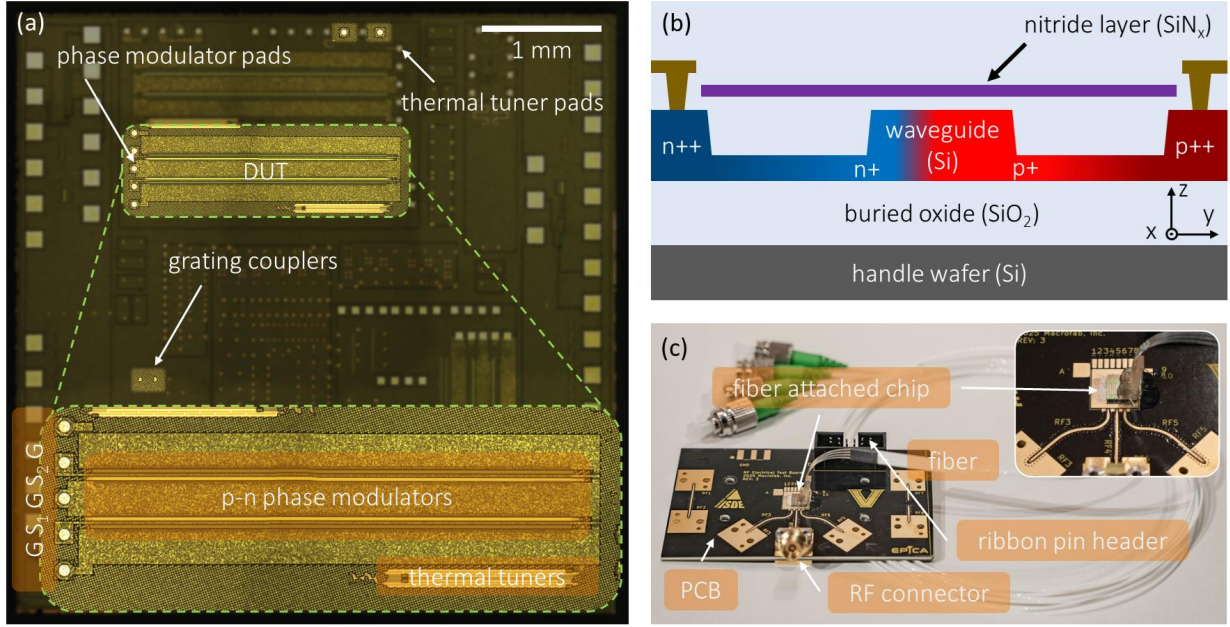

**Fig 1**. Chip and packaging details. (a) Optical micrograph of the PIC. Grating couplers, thermal phase tuner (heater) & pads, p-n phase modulator & pads, and device under test (DUT) are indicated. (b) Cross-section of the optical material stack in the GlobalFoundries Fotonix™ technology.[46] (c) Photograph of a representative test package, with a bonded and fiber attached chip mounted to the RF PCB described in the **Methods** section. The inset shows a magnified view of the fiber attach, bonded chip, and PCB.

**Total ionizing dose effects on analog and digital radio frequency performance**

To evaluate the impact of ionizing radiation on the SiPh MZM, 10-keV X-ray irradiations were performed, with device characteristics monitored between irradiation dose steps. Detailed descriptions of the radiation test setup, X-ray test conditions, and experimental electro-optic measurement setups are provided in the **Methods** section. Analog RF frequency response ($S_{21}$) and reflections ($S_{11}$) were studied using a 50 GHz two port Keysight vector network analyzer (VNA) and calibrated 70 GHz Keysight lightwave analyzer. We note that no changes in $S_{11}$ were identified so we focus here on $S_{21}$. The analog RF response of the MZM as a function of X-ray dose is shown in **Fig. 2,** where the impact of both grounded (**Fig. 2a**) and active (**Fig. 2b**) test conditions during irradiation were studied. For the grounded test, all leads are grounded during irradiation and there is no laser input to the device. For the active test, standard input conditions of -2 V bias, 0 dBm RF input (continually swept 10 MHz to 50 GHz) to the carrier depletion phase modulator in the signal arm of the MZM, quadrature bias to the thermal tuner, and 0 dBm 1310 nm continuous wave laser input to the PIC grating coupler are used during irradiation. Both tests were performed on the same sample (grounded test followed by active test) using a GGB Industries Inc. 50 Ω RF probe. We note that the RF loss added by the probe is embedded in the measurements but does not impact our analysis since repeatable measurement techniques and mitigation of sources of uncertainty are paramount to extract *changes* to device performance in reliability physics studies. Moreover, the 500 MHz – 40 GHz bias tee in the RF input path impacts the low frequency (<500 MHz). Hence, we report normalized frequency response as opposed to absolute electro-optic response. A detailed description of the normalization procedure is provided in the **Methods** section.

As shown in **Fig. 2a**, the analog RF electro-optic bandwidth of the MZM is unchanged when the sample is irradiated while grounded. In contrast, the bandwidth of the modulator decreases as a function of TID when standard active operating test conditions are used during irradiation (**Fig. 2b**). Especially at large TID (noticeable at 500 krad($SiO_2$) or more), a swift roll-off is observed up to approximately 5 GHz, followed by a plateau or a gradual tapering of the increase in degradation. The influence of charge trapping and degradation mechanisms on these trends is discussed in the **Radiation Degradation Mechanisms** section. In total, the 3 dB bandwidth decreases by nearly 10x, which is achieved before 1 Mrad($SiO_2$), indicating that the influence of TID on MZM behavior approaches saturation. We note that for all $S_{21}$ measurements, the device response is more than 20 dB above the VNA operational noise floor.

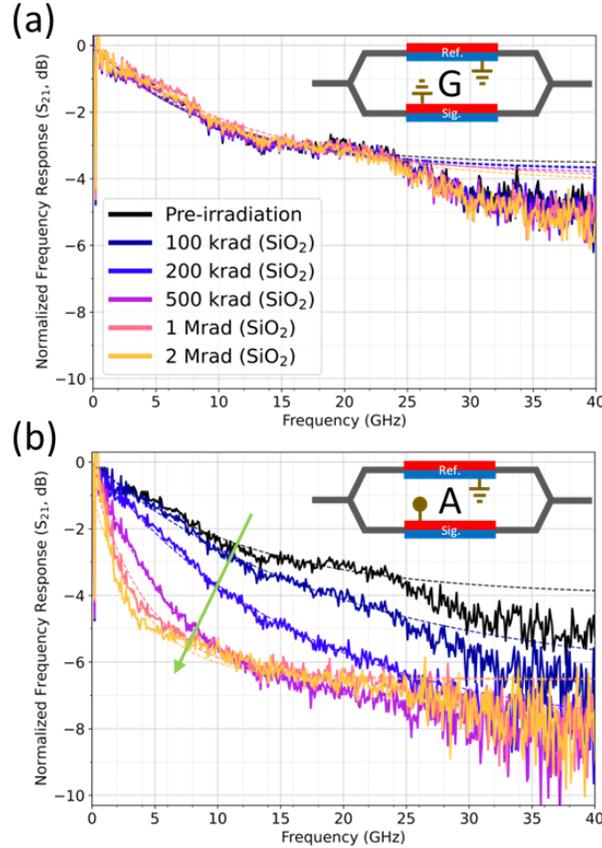

**Fig 2**. Effects of TID on frequency response of probed MZM. The normalized RF electro-optic $S_{21}$ characteristics following each X-ray dose when (a) both MZM phase shifting arms are electrically shorted and grounded during irradiation and (b) a reference phase shifting arm is grounded and the signal phase shifting arm is actively biased and monitored during irradiation. The insets further detail the conditions during irradiation (G = shorted & grounded, A = active). Active test conditions include DC bias at quadrature on the thermal tuner and -2 V with 0 dBm RF input on the high-speed phase modulator. The data from 500 MHz to 25 GHz was fit to a single exponential (dashed traces) and subsequently normalized as described in the **Methods** section.

To more clearly visualize the changes in the RF bandwidth of the MZM due to X-ray irradiation, we normalized the electro-optic frequency response at each dose step to the pre-irradiation measurement, as shown in **Fig. 3a, b**. The experiment was then repeated on a separate chip (**Fig. 3c, d**) that was packaged using wire bonding for all chip contacts as described in the **Device Design** section (**Fig. 1c**). The exponential fitting approach described in the **Methods** section is used prior to normalization of response at each dose. We note that the test sequence was also swapped during the packaged test, such that active test conditions were used for the first set of X-ray exposures followed by fully grounded test conditions. The grounded and active testing approaches were kept consistent between tests, therefore offering a comparison between both the test packaging approaches (evaluating potential for sources of uncertainty from probing within the X-ray irradiator) and the test sequence approach.

The susceptibility of the electro-optic bandwidth of the SiPh MZM to X-ray irradiation is consistent for both tests, with the frequency response relatively unchanged when the MZM is grounded during irradiation and degrading when the device is biased. The observed swift roll-off (up to 5 GHz) followed by a taper-off plateau in degradation of response is clear above 500 krad($SiO_2$), and consistent between tests. We note that slight discrepancies (e.g., modest ~1 dB electro-optic response increase in the grounded test at higher frequencies and slight fluctuation of the degradation trends in the active test) are partially attributable to the difficulty of de-embedding the RF electrical response of the PCB and normalizing the data using the exponential decay fit approach (noise at < 0.5 GHz).

The high-speed electro-optic characteristics were monitored periodically following irradiation experiments after the active irradiation tests, during which the device was left in-place within the X-ray irradiation enclosure at room temperature with all leads grounded. Partial recovery (~2 dB) of device bandwidth degradation was noted 13 hours after the completion of irradiation (green traces in **Fig. 3b, d**), with no continued recovery afterward when measured three days later, indicating all unassisted trap annealing that will occur is complete within the 12+ hours between experiments, and likely quicker. TID-induced trap recovery dynamics fall outside the scope of this work but are an appropriate direction of future study.

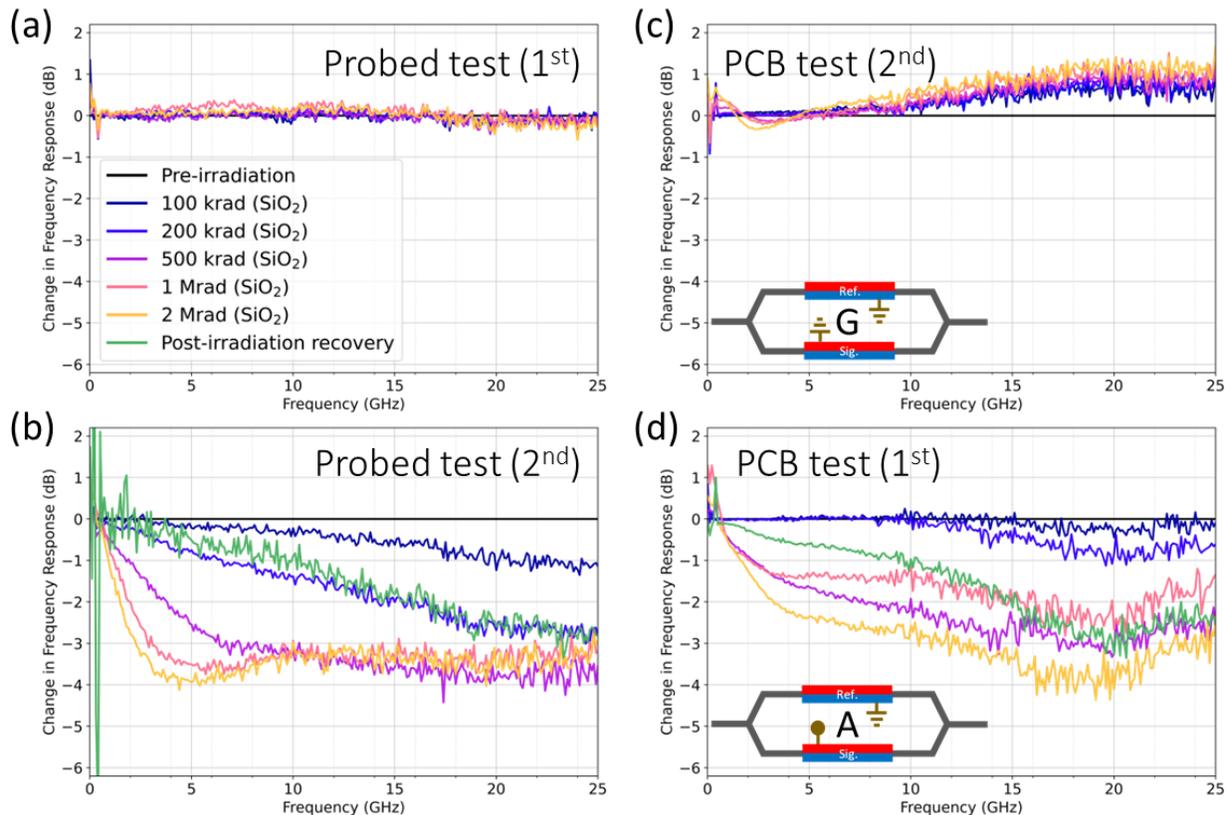

**Fig 3**. Change in electro-optic frequency response due to increasing TID on (a, b) a probed MZM and (c, d) a PCB packaged MZM. The probed MZM underwent grounded test conditions first (a), then active test conditions (b) during irradiation as described in **Fig 2**. The packaged MZM underwent active first (d), then grounded (c). Recovery (green trace) was measured 75 hours (b) and 13 hours (d) post-irradiation. The traces are limited to 25 GHz due to electrical bandwidth of the RF PCB.

The high-speed digital electro-optic characteristics were also studied, using the characterization setup detailed in the **Methods** section. Active irradiation conditions and the fully packaged approach were employed to evaluate the response given a pseudorandom binary sequence. As shown in **Fig. 4**, eye diagrams were generated using two-minute captures before irradiation (**Fig. 4a, b**), after 200 krad(SiO$_2$) (**Fig. 4c, d**), after 2 Mrad(SiO$_2$) (**Fig. 4e, f**), and 70 hours after the completion of irradiation (**Fig. 4g, h**). For the two data input rates investigated (1 GHz, 5 GHz), a discernable decrease in link fidelity is identified by slight closure of the eye diagram and a reduction in the measured peak-to-peak voltage between the one and zero level with increasing TID. Moreover, 70 hours after irradiation the eye slightly re-opens, which aligns with those observations made using the VNA and analog RF response and discussed above. The estimated bit error rates (BERs) at 1 GHz were $3.2 \times 10^{-4}$ (pre-irradiation), $5.9 \times 10^{-4}$ (200 krad(SiO$_2$)), $10.1 \times 10^{-4}$ (2 Mrad(SiO$_2$)), and $7.8 \times 10^{-4}$ (post-irradiation recovery), demonstrating an approximately three times increase in the error rate during irradiation and partial recovery after room temperature, grounded annealing (~2.5x increase from pre-irradiation BER). This recovery aligns in magnitude with those characteristics identified from the analog measurements presented in **Fig. 3**. We note that the digital experiment was performed on a separate chip from the analog experiments.

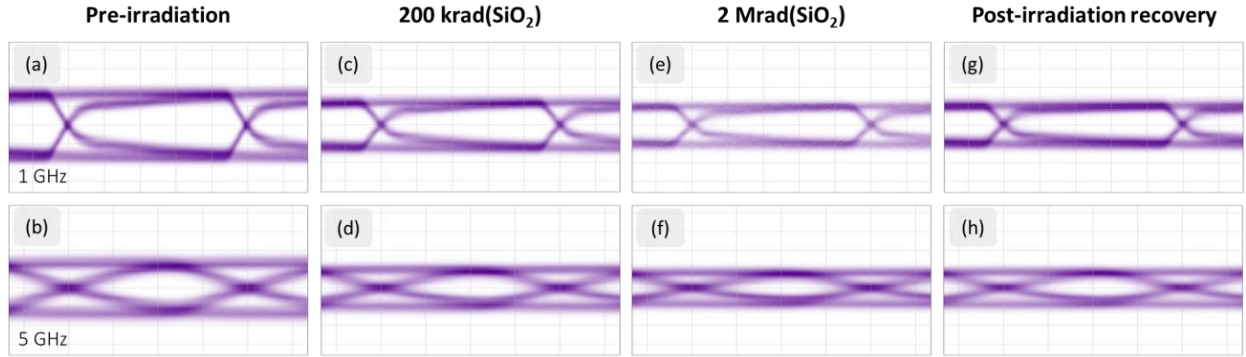

**Fig 4**. TID-induced performance degradation of the eye diagram for two baud rates. Eye diagrams are presented (a, b) before testing, (c, d) following 200 krad(SiO$_2$), (e, f) following 2 Mrad(SiO$_2$), and (g, h) 70 hours after irradiation using the fully packaged approach and active operating conditions (see **Fig. 3d**). The (a, c, e, g) 1 GHz and (b, d, f, h) 5 GHz eye diagrams are reported, as labelled in (a, b). (a-h) All plots employ the same vertical axis, with 5 mV per division (y-axis). The x-axis scales are as follows: 200 ps per division (1 GHz), and 50 ps per division (5 GHz).

**Influence of ionization on carrier depletion phase modulator characteristics**

To further investigate impacts to device characteristics in the SiPh MZM resulting from TID, we turn our attention to the carrier depletion p-n phase modulators (PMs) comprising each arm of the MZM, whose carrier dynamics are responsible for efficient modulation at GHz speeds. To avoid breakdown, we do not directly measure the PM's $V_\pi$ in reverse bias. Additionally, since the MZM layout is symmetric ($\Delta\varphi \sim 0$), there are no fringes in the optical output spectrum that would yield $V_\pi$.[47] Instead, we forward bias the PM and leverage strong carrier injection characteristics to examine phase sensitivities affecting the MZM. This DC optical characterization is performed using the setup configuration described in **Fig. 9b** in the **Methods** section. PM analysis is reported for the same device that is presented in **Fig. 3c, d**, and is shown in **Fig. 5** where $\pi$ phase cycles (contained within the attenuation envelope from optical absorption as forward bias increases) are observable. The PM character changes due to X-ray irradiation during both the grounded (**Fig. 5a**) and active (**Fig. 5b**) irradiation tests, as evidenced by the increase in bias required for one $\pi$ phase shift at increasing TID. The voltage difference of the PM to achieve a $\pi$ phase difference represents the accumulation mode PM $V_\pi$ in forward bias, which we refer to as $V_\pi^*$ (~10s of mV) to distinguish this metric from the traditional $V_\pi$ (~8V) used in the $V_\pi L$ figure of merit.

A summary of our analysis of the PM is presented in **Fig. 5c** (grounded test) and **Fig. 5d** (active test), where the estimated $V_\pi^*$ is plotted vs. TID. We fit a chirped asymmetric damped oscillator model to extract the voltage difference between inflection points and hence $V_\pi^*$ (details provided in the **Methods** section). The PM's $V_\pi^*$ consistently increases with TID and the rate of increase is approximately 50% greater during the active irradiation test (**Fig. 5d**). This characteristic was similarly observed for the experiment when grounded irradiation was performed first in the test sequence (data not shown for redundancy), adding confidence to the earlier observation that the results are agnostic to the test sequence.

During the grounded irradiation test sequence, the signal and reference arm PMs are both grounded such that PM changes are balanced (within inherent device-to-device variability). Conversely, during the active irradiation test sequence, the signal arm PM is biased (-2 V + 0 dBm 10 MHz to 50 GHz inputs) while the reference arm PM is grounded, establishing the premise of imbalance in radiation effects between the signal and reference arms of the SiPh MZM when irradiated under standard operating conditions. With this understanding, we now consider the power output trends near the standard PM operating bias (-2 V), which are indicated by the green arrows in the insets of **Fig. 5a, b** where the full sampled voltage range is shown. Since the PM phase sensitivity is lower in reverse bias (compared to forward bias carrier injection), the output power becomes a function of the interference (intrinsic phase) and attenuation difference between the two arms, which are both affected by irradiation (and affected unequally during the active test). Near the PM operating bias, the trend in output power change reverses with increasing TID; this observation similarly persists in the MZM transfer function ($P_{opt.}$ vs. $V_{TPT}$), where changes to peak (DC) optical power output and bias point for quadrature operation reverse around 200-500 krad($SiO_2$). For the grounded and active experiments, the quadrature bias point changes by up to 0.2 V, but the TPT $V_\pi$ extracted from the thermal tuner transfer function does not change. Hence, while changes to the intrinsic phase difference between the signal and reference arm PMs are accounted for by updating the quadrature bias point, the efficiency of phase modulation is not, leading to the RF performance degradation seen in the analog and digital characteristics shown above (**Figs. 2-4**).

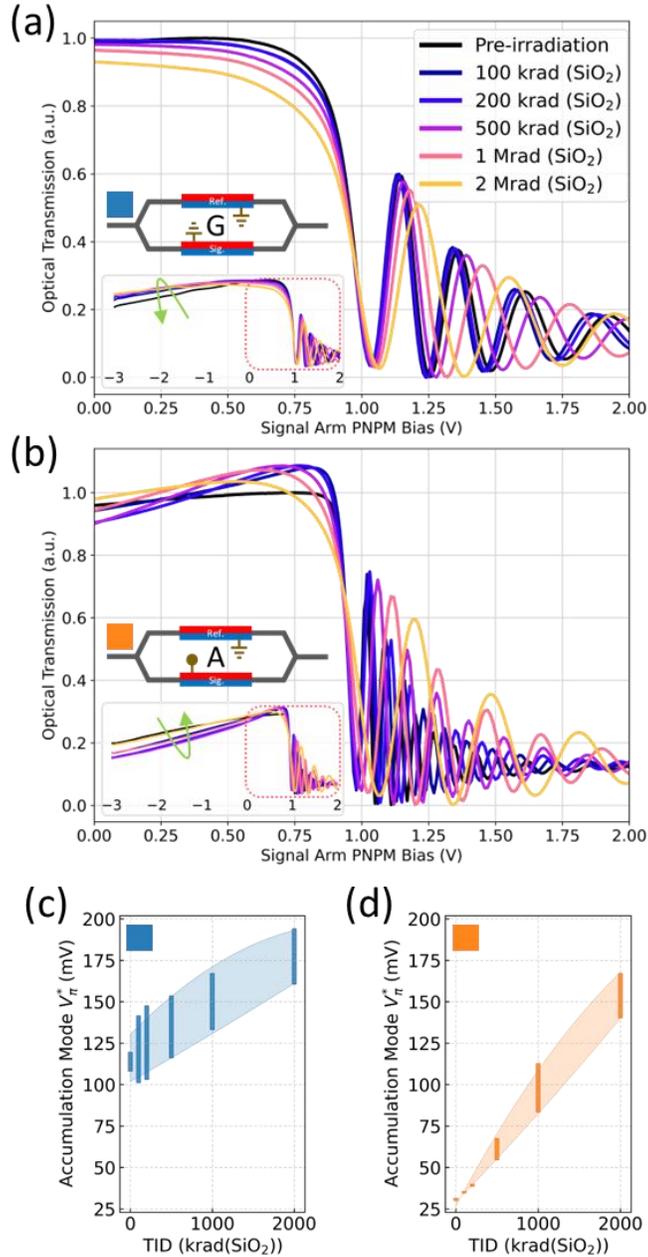

**Fig 5**. p-n phase modulator (PM) accumulation mode characteristics under TID. (a, b) The MZM optical power output as a function of forward bias applied to the PM for (a) grounded and (b) active conditions during irradiation. Full traces (-3 to 2 V) are shown inset, with reverse bias power output trends with TID indicated by the green arrows. The inset illustrations indicate the grounded and active conditions employed during irradiation (see **Fig. 2** caption). (c, d) PM $V_\pi^*$ estimated from a damped, chirped oscillator fit of the data in (a, b) plotted for (c) grounded and (d) active irradiation conditions. The span (y-direction) of the markers represents the range of $V_\pi^*$ extracted between 1 and 2 V. The range between the maximum and minimum $V_\pi^*$ was fit to a quadratic polynomial to guide the eye (shaded region). Color markers denote paired results (a, c) blue = grounded conditions, (b, d) orange = active conditions.

**Radiation degradation mechanisms**

To further examine the physical mechanisms leading to TID-induced performance degradation in the SiPh MZM, a two-dimensional (2D) TCAD model of the PM waveguide cross-section was implemented using Synopsys Sentaurus Device. A detailed description of the TCAD simulation parameters is provided in the **Methods** section. The net difference between electrons and holes, or space charge profile, in the doped silicon PM is calculated at -2 V for the pristine device (**Fig. 6a**) and after modelling TID exposure (**Fig. 6b**). To model the impact of TID in the oxide, a volume charge distribution is projected onto the semiconductor/oxide interface, representing the effects of net hole trapping in the oxide.[38,39] Accordingly, sheets of positive interface charge representing trapped holes were placed at all Si/SiO$_2$ interfaces, with an areal density of $N_{ot} = 10^{12} cm^{-2}$,[48] as indicated by the red "+" labels in **Fig. 6b**. The third dimension is modelled as infinitely extended, so carrier concentration is uniform along the length of the PM waveguide. Upon hole trapping in the oxide, negative charges within the silicon waveguide layer attempt to neutralize the positively charged SiO$_2$ by moving toward the Si/SiO$_2$ interfaces; we note that this inversion layer depletion was also observed by Zeiler, *et al.*[12]. Focusing on the p-region within the waveguide ridge, the region where high concentrations (~10$^{18}$ cm$^{-3}$) of electrons are present increases by more than 20 nm from the waveguide edges toward the waveguide center. As shown in **Fig. 6c**, the fundamental transverse electric (TE) optical mode overlaps strongly with the outer waveguide area, especially toward its top and bottom surface. Upon application of bias (both DC and AC), these positive trapped charges also act as a potential barrier, preventing the flow of charge carriers out of the waveguide toward the anode contact. Decreased charge flow results in a decrease in the contrast in refractive index (and subsequent optical phase accumulation difference) between the on and off modulator states,[49,50] which explains the degradation in modulator efficiency observed in the PM (**Fig. 5**) and a linear decrease in the RF response characteristics (**Figs. 2-4**).

While this explanation is sufficient up to ~200 krad(SiO$_2$), the nature of trap activation must also be considered to explain the frequency dependence of the electro-optic degradation (swift roll-off up to ~5 GHz for TID ≥ 500krad(SiO$_2$) as in **Figs. 2-3**). So-called "fast traps", which have capture and emission time constants on nanosecond time scales, likely form close to the semiconductor conduction band edge. Fast traps have been identified as the source of frequency-dependent TID effects on electronics such as gate lag, RF power gain, and increased channel resistance and device capacitance in high-electron mobility transistors (HEMTs) and silicon-on-insulator field-effect transistors (SOI FETs).[51–53] While the DC characteristics of the PM are length-dominated and mask any changes to *I-V* or *P-V* characteristics (dark current on mA scale), the rate that these interface traps are filled and vacated aligns with the response in the GHz range to affect the MZM frequency response.

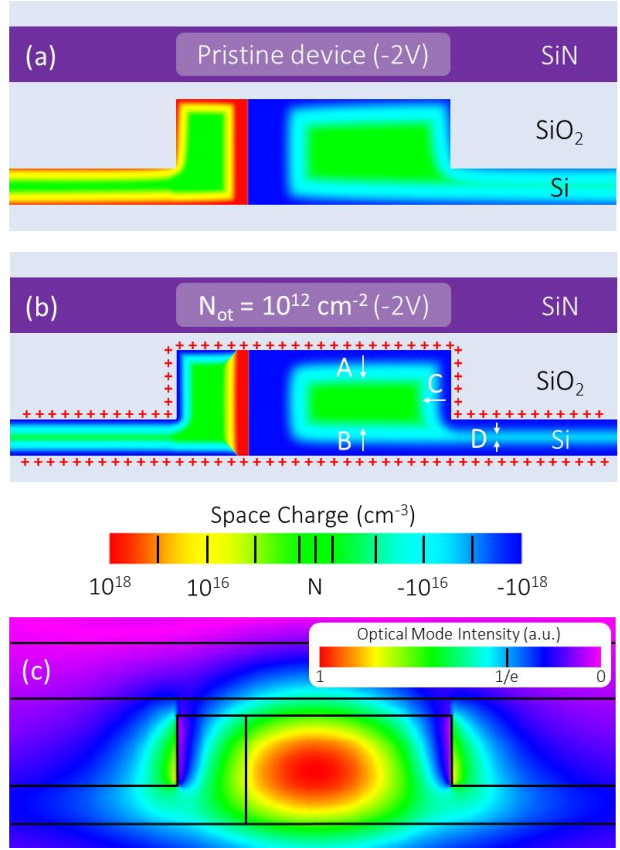

**Fig 6**. Space charge distributions. (a, b) The net difference of electrons and holes (space charge) for a 2D cross-section of the PM waveguide at -2 V (a) before irradiation (pristine device) and (b) after irradiation. (b) $10^{12}$ cm$^{-2}$ positively charged sheets are modelled at each Si/SiO$_2$ interface and indicated by red "+" markers, representing the formation of oxide charge traps ($N_{ot}$). The A, B, C, D labels accompanied by white arrows indicate the areas within the waveguide where electrons flow at high concentrations toward the positively charged interface. (a, b) The color scale bar is logarithmic with markers at each order of magnitude and net hole density and net electron density represented by positive and negative values, respectively. The Si, SiO$_2$, and SiN layers in the model are labelled. (c) The normalized TE optical mode profile in the waveguide is shown. The 1/e point is indicated in the color bar scale.

Finally, we consider the locations along the waveguide's Si/SiO$_2$ interfaces where traps are most likely to form and fill in the largest concentrations to most significantly contribute to the experimentally observed TID-induced MZM performance degradation. We note that charge trapping likely occurs at all locations along the Si/SiO$_2$ interfaces, but at swifter rates and higher densities with increasing local electric field magnitudes.[12] **Fig. 7** shows electric field profiles for the PM waveguide under no applied bias (**Fig. 7a**) and -2 V (**Fig. 7b**). The maximum electric field magnitude increases by approximately two-times when -2 V is applied ($|\vec{E}_{max}|$ is 4×10$^5$ V/cm at 0 V in **Fig. 7a** vs. 8×10$^5$ V/cm at -2 V in **Fig. 7b**). Additionally, nearly the entire perimeter along the top and bottom ridge waveguide interface experiences electric field magnitudes greater than 10$^4$ V/cm, which is around the magnitude at which trap formation accelerates.[54–56] Hence, regions A and B in **Fig. 6b** are likely to be the most consequential to the TID effects we report. The electric field magnitude along y- and z-axis cuts at the waveguide Si/SiO$_2$ interfaces and through the waveguide center are also shown (**Fig. 7c**), illustrating the spatial distribution of the fields relative to the device interfaces for both bias conditions. We note that when -2 V is applied, the electric field magnitude also exceeds 10$^4$ at the SiN/SiO$_2$ interface. Hole trapping at this location may mildly affect the static phase difference (compensated for by adjusting the quadrature bias point) but are unlikely to significantly impact phase modulation in the silicon layer. The simulated difference in electric field magnitude explains the experimentally observed TID sensitivity dependence on irradiation test conditions. When all leads are grounded and the field is low, fewer traps form. When the device is actively biased during irradiation, trap formation increases, and the device performance degrades significantly.

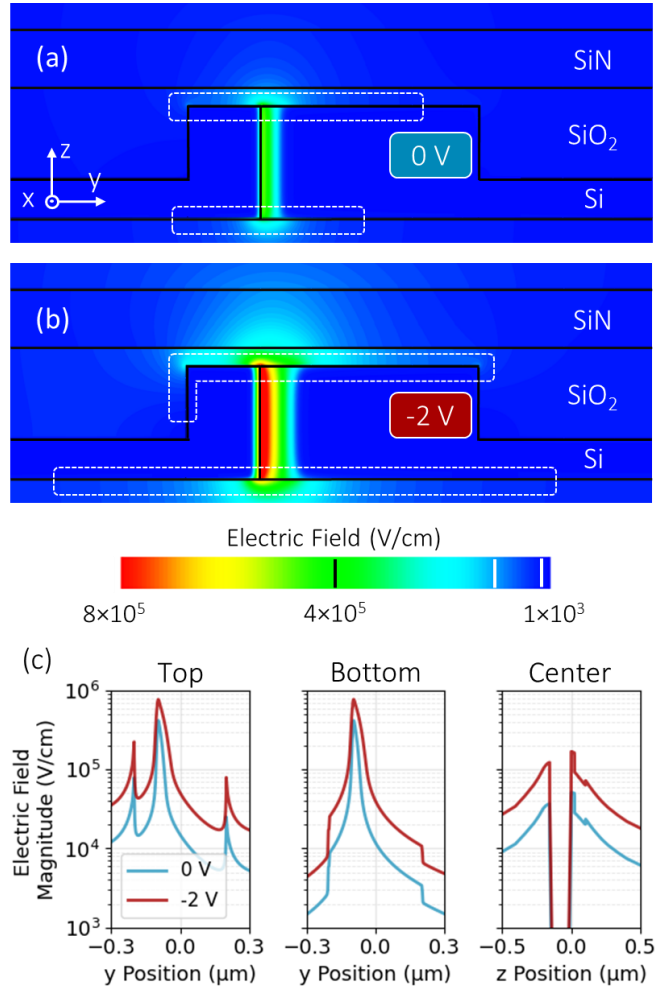

**Fig 7**. Electric field profiles. (a, b) The electric field magnitude for a 2D cross-section of the PM waveguide at (a) 0 V and (b) -2 V. (a, b) The color scale is linear with a minimum limit of $10^3$ V/cm. The white dashed boxes indicate regions along the $Si/SiO_2$ interface where the electric field magnitude is $10^4$ V/cm or greater. The white markers without a text label in the color bar scale indicate $10^5$ and $10^4$ V/cm, respectively. The Si, $SiO_2$, and SiN layers in the model are labelled. Biasing results in greater electric field magnitudes which increase trap formation and occupation during irradiation. (c) Electric field magnitude along (left) the y-axis at the top $Si/SiO_2$ waveguide interface, (middle) the y-axis at the bottom $Si/SiO_2$ interface, and (right) the z-axis cut through the center of the waveguide at 0 V (blue) and -2 V (red). The y position of 0 μm corresponds to the waveguide center and z position of 0 μm corresponds to the top $Si/SiO_2$ interface. Discontinuities are attributed to the $Si/SiO_2$ and $SiN/SiO_2$ interfaces and corners.

**Testing requirements for photonic devices and circuits in harsh radiation environments**

The testing approach selected for space-mission radiation effects qualification plays a large role in the susceptibility identified within devices. Hence, choosing testing conditions that best mimic those when the devices are operating in their deployed environment is essential, acknowledging that ground-based radiation testing often requires accelerated exposure, sources of a single radiation species and energy at a time, and analogs to the types of radiation that are present in the space environment.[57–60] The limitations in ground-based testing have led to substantial interest in on-orbit testing of integrated photonics chips, with Mao, *et al.*[43] and Hoff, *et al.*[44] each showing characterization of photonic chips before and after space flight on the international space station in lower Earth orbit (LEO). For these first opportunities to examine photonics in space, no data was taken during space flight and the samples were attached to a larger payload and likely electrically floating during exposure. **Fig. 8** demonstrates why this approach is not sufficient for fully characterizing MZMs for operation in a space environment, with analog electro-optic frequency response presented for an irradiation test on a SiPh MZM sample using the probed measurement approach, but with the reference arm PM left floating. Comparing this result to the that reported in **Fig. 2a**, notable degradation is observed with increasing TID when one PM arm is floating during testing, which is likely due to static charge accumulation in sensitive active regions in absence of a dissipation pathway during irradiation. Uncontrolled releases of built-up charge can cause electro-static damage and other forms of device degradation. We note that the slight dive and peak in the characteristics seen in the 100 and 200 krad($SiO_2$) measurements below 10 GHz in **Fig. 8** is attributed to a slight shift in the contact alignment of the RF GSG probe to the device padset. While the dose required to noticeably damage the part under 10-keV X-ray irradiation occurs at a very large TID threshold (200 – 500 krad($SiO_2$)) compared to the dose levels achieved for relatively short exposures (< 1 year), in low Earth orbit (< 15 rad), the actual space environment also includes extremely energetic, charged particles (e.g., protons, α-particles, heavy ions) that can cause charge collection in sensitive regions when no pathway for dissipation is provided. Thus, a particle strike that would cause a SET when contacted might instead cause static charge buildup or damage when the device is floating. These influences work against interpretation of results from samples which are flown in space but not properly contacted. Moreover, probing, especially when performed months or years apart, provides extra opportunities for sources of error.

The work presented here emphasizes the desirability of additional resource allocation to integrated photonic devices and circuits included in space flight systems, especially in leveraging standard operating conditions for active devices and collection of on-orbit data when possible. As more opportunities and resources become available, the knowledge gained from these ground-based testing results can help inform appropriate deployment requirements and experimental strategies for efficiently collecting useful data.

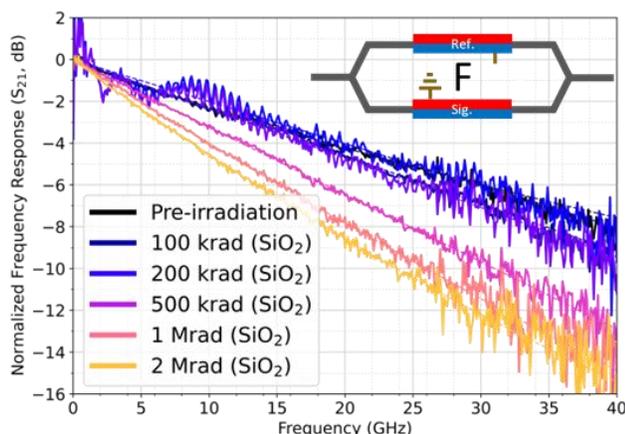

**Fig 8**. Floating contacts during irradiation. The normalized RF electro-optic $S_{21}$ (dB) characteristics are plotted following each X-ray dose step when the MZM signal arm is probed and grounded during irradiation (as in **Fig. 2a**), but the reference phase shifting arm is left floating. The inset illustration indicates the conditions during irradiation. Linear fits used to normalize the 0 GHz DC insertion loss for each dataset are plotted (dashed traces).

Since the trends observed in this work are noted largely irrespective of the test sequence employed (grounded first vs. active first) and the packaging approach (probed vs. RF PCB), we confirm that both approaches are suitable for characterizing radiation effects in photonic devices and circuits for ground-based *in situ* testing. Key trade-offs to consider for radiation and other reliability testing approaches on photonics include (1) probed testing minimizes loss, which is important when performing RF testing and identifying frequency dependent loss from test equipment, experimental setups, and fiber-to-chip coupling. Comparatively, the RF PCB design used for the SiPh MZM studies presented in this work is sufficient, but future improvements to the design could mitigate characterization uncertainty and improve comparative quantification. (2) While probing is appropriate for X-ray testing in ARACOR irradiators, a stereomicroscope is generally needed, which is prohibitively difficult to implement at test facilities where vacuum systems are employed for radiation effects testing, the beam line is parallel to the floor, and facility rates are expensive ($1000's/hour). In these cases, the setup complexity and time consumed by preparing a microscope for dynamically aligned probes makes this approach inadvisable. Conversely, a fully packaged sample can be employed with cable feedthroughs for effective setup and testing approaches.

**Conclusion**

Significant total ionizing dose effects in silicon photonic MZMs actively operated under standard conditions during 10-keV X-ray irradiation are measured, with significant deterioration to the high-speed electro-optic frequency response, closure of the eye diagram, and frequency dependent increases in the estimated error rate observed. Conversely, irradiation while the MZMs are shorted and grounded causes no device performance degradation and therefore fails to identify the high-speed electro-optic susceptibility of the devices when operating in a space environment. Moreover, irradiation while the reference modulator arm is floating introduces damage that is not related to realistic degradation modes when the device is contacted within a fully functional deployed system in space. These results emphasize the importance of test conditions in qualifying photonic parts for space missions and elucidating physical degradation modes in harsh environments. Underlying physical degradation mechanisms are identified using experiment and TCAD simulations, where negative charges flow toward the positively charged interface as oxide traps form and collect in high concentrations toward the waveguide edges in the p-region of the PM waveguide. Modulation efficiency reductions in the PM result, especially when a signal arm is biased during irradiation. Moreover, the frequency dependent deterioration of the RF bandwidth is attributed to the formation of "fast traps," which do not impact DC characterization. These traps capture and emit holes at energies close to the conduction band edge causing swift roll-off in the electro-optic response at intermediate frequencies (up to ~5 GHz). In a broad context, the performance degradation of MZMs under X-ray irradiation identified in this work are modest and do not preclude the use of similar MZMs in space missions; however, as performance requirements and circuit complexity of integrated photonic technologies continue to increase, these results will become critical to PIC design considerations for operation in space, accelerator, and nuclear defense environments.

**Acknowledgement**
This work was supported by the National Science Foundation (NSF) Industry-University Cooperative Research Center (IUCRC) Electronic-Photonic Integrated Circuits for Aerospace (EPICA) award #2052742 and the NSF EPICA industry advisory board (IAB). KPA's effort in this work was supported by the National Aeronautics and Space Administration (NASA) Space Technology Graduate Research Opportunities (NSTGRO) 2023 fellowship award #80NSSC23K1198. NJK's effort in this work was supported by the Vanderbilt University NSF Establishing Multimessenger Astronomy Inclusive Training (EMIT) Program #2125764. Funding for equipment was supported in part by the Air Force Office of Scientific Research (AFOSR) Defense University Research Instrumentation Program (DURIP) award

#FA9550-22-1-0471. John Fellenstein of the Vanderbilt Arts and Science Machine Shop (Nashville, TN, USA) is acknowledged for fabrication of custom test packages, fixtures, and mechanical experimental setup components. Scott Ferguson and Jeffrey Murphy of Keysight Technologies (Santa Rosa, CA, USA) are acknowledged for assistance in equipment calibration, troubleshooting, and technical discussions. Steven Kosier and James Trippe of Vanderbilt University (Nashville, TN, USA) and Solomon Musibau of imec (Leuven, BE) are thanked for useful technical discussions. Mark Stephen, Fabrizio Gambini, Victor Torres, and Landen Ryder of NASA Goddard Space Flight Center (Greenbelt, MD, USA) are acknowledged for their mentorship and useful technical advice. Michael McCurdy and Phoenix Harris of Vanderbilt University (Nashville, TN, USA) are acknowledged for technical support.

## Methods

*Chip preparation and electrical packaging:*

Chip samples were mounted to specially designed printed circuit boards (PCBs) using fast drying silver paste (Ted Pella 16040-30). Ribbon pin headers were used for DC bias to the thermal tuner and for connecting the PCB ground plane. Chip contacts were ball-bonded using 0.7 mil gold wire. For probed experiments, a GGB Industries Inc. 50 Ω GSG RF probe with less than 2 dB loss at 50 GHz was used to address the signal arm PM ($S_1$) and the adjacent ground contacts. The reference arm PM contact ($S_2$) and remaining ground contact were bonded and shorted and grounded or biased as described. For fully packaged tests (as indicated in **Fig. 2(c)**), a 50 Ω 2.4 mm connector jack solderless PCB mount was aligned using a stereoscope and compression affixed to a co-planar trace on the PCB, which was designed in Rogers 4350B dielectric material. Here, ball bonds were necessary due to the design of the GSGSG pads, which included an oxide liner around the pad intended for easy alignment of probes that prevented successful wedge bonding.

*Optical fiber array bundle attach:*

Attachment of a 10° (matching the preferred grating coupler angle) optical fiber array with four single mode fibers having FC/APC connectors (as indicated in Fig. 1(c)) was performed using a custom-built photonics characterization setup specially designed to support precision grating coupling of a continuous-wave laser source. A Santec TSL-550 tunable laser (set to 1310 nm), Newport 818-IG-L-FC/DB fiber coupled photodiode detector, and Keithley 2400 SMU were used for free space fiber alignment and optimization during fiber attachment. Thorlabs piezo inertia actuators, Thorlabs Kinesis® K-Cube™ piezo inertia actuator controllers, and manual polarization selection paddles were used to position the fiber and select transverse electric (TE) polarization, for which the grating coupler is designed. Fiber attachment was performed using a specially formulated UV-curable optical epoxy (Masterbond UV10TKLO-2). The fiber attachment procedure is described in detail by Arnold.[61] Additionally, the radiation response of the entire fiber attach assembly (short ~2 in. fiber segment, fiber array block, cured epoxy, grating couplers connected to plain silicon waveguide) was studied under 10-keV X-ray irradiation, showing robustness of the fiber attach process for all total ionizing doses considered in this work.[62]

*Experimental setups:*

Device characterization was performed using the experimental setups described in **Fig. 9**. For all measurements, the continuous-wave (CW) Santec TSL-550 tunable laser was set to 1310 nm and routed through single mode FC/APC fibers and manual polarization paddles to the fiber bundle attached to the input and output grating couplers. Single mode FC/APC fibers (optical, blue), Bayonet Neill-Concelman (BNC) cables (DC electrical, green), and 1 meter 50 GHz 2.4 mm W. L. Gore & Associates cables with 2.4 mm (50 GHz) and 1.85 mm (65 GHz) adaptors (RF electrical, red) were used for all experiments. Keithley 2400 Source measure units (SMUs) were used to apply bias and perform voltage sweeps on the thermal tuner and signal arm PM. For electrical characterization (phase modulator arm *I-V* characteristics), the laser input was shuttered. For all DC optical characterization (*P-V* of thermal tuner and signal arm PM), a Newport 818-IG-L-FC/DB fiber coupled photodiode detector was used in tandem with an additional

Keithley 2400 SMU to measure the detector photocurrent.

High-speed electro-optic measurements were performed using a 50 GHz two port Keysight N5245B PNA-X vector network analyzer (VNA), which can measure electrical frequency response 10 MHz – 50 GHz. A Picosecond Pulse Labs 500 MHz – 40 GHz, 100 mA bias tee was connected to the GSG probe (probed measurement) or PCB jack connector (bonded measurement), simultaneously connecting port 1 of the VNA (RF input) and a Keithley 2400 SMU (DC input) to the signal arm PM. The single mode FC/APC patch fiber output from the MZM was connected to a calibrated N4377A 70 GHz lightwave detector, which was in turn connected to port 2 of the VNA and de-embedded from the analog RF characteristics. All equipment used is designed for 50 Ω impedance. For probed measurements, the bias tee has the smallest electrical bandwidth of all components, so RF characterization up to 40 GHz could be considered. For bonded measurements, the combined optical loss (determined by the accuracy of fiber alignment after attach is performed) and RF PCB loss determined the electro-optic bandwidth that could be considered, which limited measurements to around 25 GHz.

Digital electro-optic measurements were performed using an Anritsu MP1763C 0.05 – 12.5 GHz pulse pattern generator with a pseudorandom binary sequence (250 mV peak-to-peak output). An Agilent E8257D 250 kHz – 31.8 GHz analog signal generator was used to clock the pulse pattern generator (0 dBm RF electrical power output). The digital stream was coupled to the MZM using the Picosecond pulse labs bias tee, as in the analog test sequence above. The single mode FC/APC patch fiber output from the MZM was connected to an uncalibrated New Focus 1414 25 GHz fiber optic detector, where the digitally modulated output was converted to the electrical domain and amplified using a SHF Communication Technologies 100 AP 40 KHz – 25 GHz RF amplifier. The RF amplifier was biased at +9 V using a Keithley 2230-60-3 triple-channel DC power supply to achieve a gain of approximately 15 dB. Finally, the amplified electrical signal was measured using a Teledyne-Lecroy Labmaster 10 Zi-A 36 GHz oscilloscope. The eye-diagram of the pseudorandom binary sequence was recorded using 80 G samples/s for a two-minute interval and the bit error rate was estimated using the internal Teledyne-Lecroy Labmaster oscilloscope software.

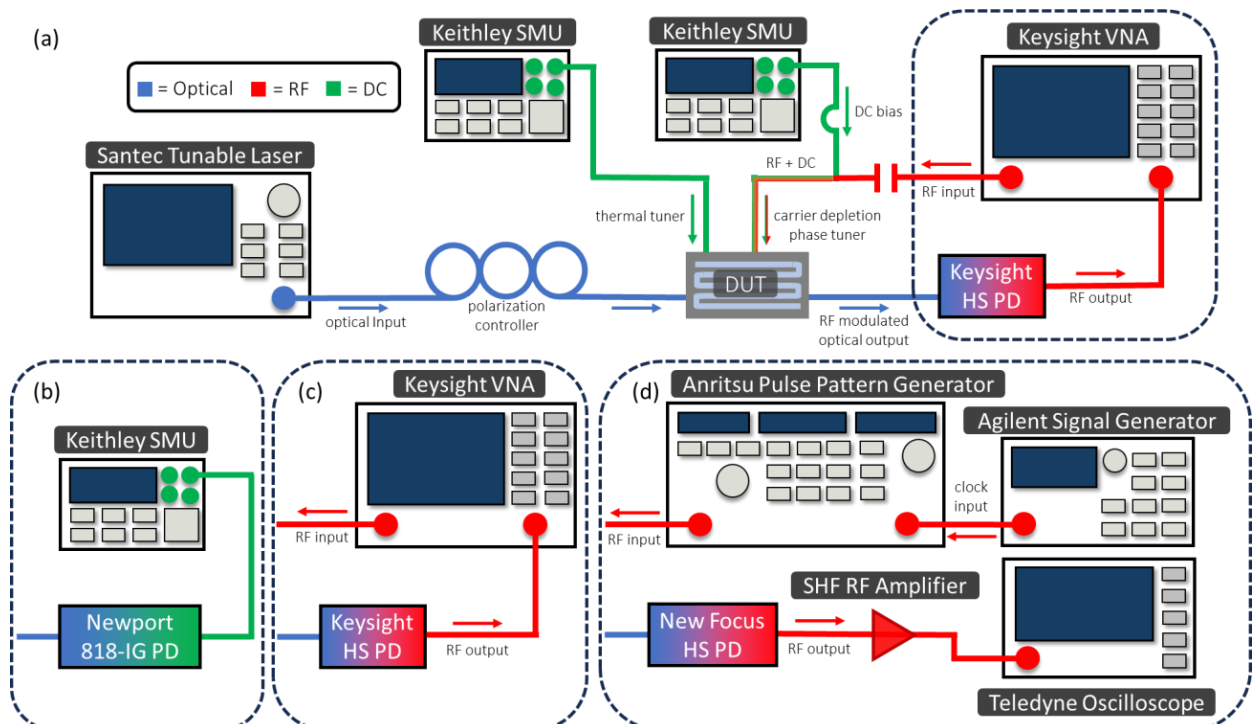

**Fig. 9**. Experimental setup details. (a) Diagram of optical (blue), DC electrical (green), and RF (red) equipment used to control and characterize the MZM. Fibers and electrical cables are swapped to connect equipment as outlined in the dashed boxes for (b) alignment prior to fiber attachment and DC optical

characterization, (c) analog RF electro-optic characterization (also shown in (a)), and (d) digital electro-optic characterization.

*Radiation test setup and X-ray test conditions:*
Total ionizing dose effects studies were performed at Vanderbilt University in Nashville, TN, using an ARACOR 4100XPL 10-keV X-ray Irradiator.[63] Measurements were performed *in situ*, but not in flux, starting three minutes after completion of radiation exposure. Hence, devices were not moved and all electrical and optical connections were fixed throughout the irradiation and measurement procedure. For all irradiations, the X-ray dose rate was approximately 28.14 krad(SiO$_2$)/min. The sample was placed 1.75 inches from the aperture of the X-ray source and leveled before the start of each test sequence, ensuring that the angle of incidence and dose rate were held constant during irradiation. For all experiments presented on the same sample (grounded and active test with one test package), at least 12 hours passed between irradiation sequences (at least overnight). This was required due to the total duration of setup, irradiation, and measurement for each sample. Optical and electrical cables were fed into the steel X-ray test chamber, which is closed and latched during irradiation, using a cable crunch. The back side of the sample, test package, and X-ray test chamber mounting stage were shorted and directly connected to Earth ground using existing electrical infrastructure in the laboratory. For all measurements after irradiation, the ambient temperature of the measurement stage was monitored; no variation greater than 2 °C was noted, which does not meaningfully affect the MZM response.

During the grounded irradiation test, all electrical pads and leads were shorted and grounded using the PCB test package connections. During the active irradiation test procedure, the VNA was swept 10 MHz to 50 GHz using 501 frequency points in intervals of approximately ½ seconds. The quadrature bias measured after each previous X-ray irradiation dose was set on the thermal tuner, although this bias to the resistive heater does not likely influence device effects from irradiation. The reference modulator arm was grounded through the PCB test package using wire bonds for all irradiation procedures. 0 dBm 1310 nm continuous-wave laser input was used during irradiation and was increased to +5 dBm during measurement following irradiation. Further, the MZM transfer function was measured after each irradiation and a new quadrature bias setpoint on the thermal phase tuner was implemented before the RF analog characteristics were collected.

*Normalization procedure for analog RF electro-optic response measurements:*
An exponential decay fit routine was used to fit all RF electro-optic bandwidth measurements following: $FR_{fit}(f) = Ae^{-bf} + IL_{DC}$, where the frequency response fit ($FR_{fit}$) is a function of frequency ($f$), ($IL_{DC}$) is the DC insertion loss at 0 GHz, and $A$ and $b$ are free-fitting variables. The 40 GHz bias tee introduces noise at frequencies less than 500 MHz, so the fit was calculated in a window between 500 MHz and 25 GHz. The $IL_{DC}$ y-intercept determined by each fit is used to normalize each trace to zero insertion loss at 0 GHz. We note that the exponential fit falls back to a linear fit if the residual of the exponential approach is large, which is necessary when the frequency response is linear as in **Fig. 8**. This normalization procedure eliminates any influence from laser power drift or fluctuation, uncertainty in the polarization alignment, or slight shifts in the RF probe contact.

*Fit procedure for carrier depletion phase modulator characteristics:*
To quantify the changes to the forward bias PM modulation efficiency shown in **Fig. 5**, a chirped asymmetric damped oscillator model is used to fit the dataset from each dose between +1 and +2 V, using:

$$P_{opt.}(V) = Ae^{-V/\tau_1} * cos\left(2\pi\left(\frac{V-V_0}{V_\pi} + \alpha(V-V_0)^2\right)\right) + De^{-V/\tau_2} + C \qquad \text{Eq 2}$$

where $A$ is the amplitude and $\tau_1$ is the decay constant for the oscillatory component, $V_\pi$ is the effective voltage swing to induce a $\pi$ phase shift, $V_0$ is the center voltage, $\alpha$ is the chirp parameter, $D$ is the amplitude and $\tau_2$ is the decay constant for the background decay, and $C$ is the y-intercept offset. This fit approach is an approximation and does not necessarily represent the exact physical phenomena observed but was necessary due to the voltage division used to measure the forward bias characteristics (0.02 V). Thus, all variables are free-fitting to closely represent the measured data and rather than use the $V_\pi$ parameter fit from the model, we use the spacing between inflection points throughout the oscillation (calculated using second derivatives) to estimate the PM $V_\pi^*$.

*TCAD simulations:*

The 2D TCAD model of the PM waveguide cross-section was implemented using Synopsys Sentaurus Device. Implant profiles were calibrated using sheet resistance measurements and *I-V* characteristics (-3 to 2 V) from the foundry PDK specifications. A Gaussian decay of each implant was implemented at the interface between the n and p regions to estimate the depletion layer. Coarse 100 nm meshing was used in the oxide bulk where the electric field intensity is low and fine 5 nm meshing was used near the $Si/SiO_2$ interfaces and within the doped silicon PM. An ultra-fine 2 nm mesh step was used in the depletion region where the electric fields are strongest. Doping-dependent Shockley-Read-Hall generation and recombination, doping-dependent carrier mobility, and Auger recombination models were used to model the physical characteristics of the device. The simulation temperature was set to 300 K and a length scaling factor (2 mm) was implemented to approximate the 3D device. Sidewall sloping was neglected in simulation and is not expected to play a significant role in charge dynamics of the PM.